\documentclass[twocolumn,showpacs,preprintnumbers,amsmath,amssymb]{revtex4}
\usepackage{graphicx}
\usepackage{bm}
\usepackage{dcolumn}

\def\sci#1#2#3{{Science} {\bf #1}, #2 (#3)}
\def\prl#1#2#3{{ Phys.   Rev.   Lett.  } {\bf #1}, #2 (#3)}
\def\pla#1#2#3{Phys.   Lett.   A {\bf #1}, #2 (#3)}
\def\pra#1#2#3{Phys.   Rev.   A {\bf #1}, #2 (#3)}
\def\pre#1#2#3{Phys.   Rev.   E {\bf #1}, #2 (#3)}
\def\jpa#1#2#3{J. Phys.   A {\bf #1}, #2 (#3)}
\def\epl#1#2#3{Europhys. Lett. {\bf #1}, #2 (#3)}

\def\jsp#1#2#3{J.   Stat.   Phys.   {\bf #1}, #2 (#3)}

\def\physa#1#2#3{Physica A {\bf #1}, #2 (#3)}

\def\noi{\noindent}
\def\bc{\begin{center}}
\def\ec{\end{center}}
 \newcommand{\bea}{\begin{equation}}
 \newcommand{\eea}{\end{equation}\noi}
 \newcommand{\ber}{\begin{eqnarray}}
 \newcommand{\eer}{\end{eqnarray}\noi}
\begin{document}
\title{More Accurate Theory for Bose-Einstein Condensation Fraction}
\author{Shyamal Biswas}\email{tpsb@iacs.res.in}
\affiliation{Department of Theoretical Physics,
Indian Association for the Cultivation of Science \\
Jadavpur,Kolkata-700032, India}
\date{\today}
\begin{abstract} 
          In the thermodynamic limit the ratio of system size to thermal de Broglie wavelength tends to infinity and the volume per particle of the system is constant. Our familiar Bose-Einstein statistics is absolutely valid in the thermodynamic limit. For finite  thermodynamical system this ratio as well as the number of particles is much greater than 1. However, according to the experimental setup of Bose-Einstein condensation of harmonically trapped Bose gas of alkali atoms this ratio near the condensation temperature($T_c$) typically is $\sim 32$ and at ultralow temperatures well below $T_c$ a large fraction of particles come down to the single particle ground state, and this ratio becomes comparable to 1. We justify the finite size as well as ultralow temperature correction to Bose-Einstein statistics. From this corrected statistics we plot condensation fraction versus temperature graph. This theoretical plot satisfies well with the experimental plot(A. Griesmaier et al..,Phys.Rev.Lett.\ \ {\bf{{94}}}{(2005){160401}}).
\end{abstract}
\maketitle         
             Let us consider a noninteracting many particle system of Bose gas containing $N$ alkali atoms. The system is in equilibrium with its surroundings at temperature($T$). The mass of a particle is $m$. The length and the volume of the system are $L$ and $V(=L^3)$ respectively.  Thermal de-Broglie wave length of a single particle is $\lambda_T=\sqrt{\frac{2\pi\hbar^{2}}{mkT}}$, where $k$ is the Boltzmann constant. Average separation of particles is $l=(\frac{V}{N})^{1/3}$. In the classical limit $\frac{l}{\lambda_T}\gg 1$, i.e. $\frac{kT}{\frac{2\pi\hbar^2}{mL^2}}\gg N^{2/3}$. At sufficiently low temperatures when $\frac{l}{\lambda_T}\sim 1$, the gas becomes degenerate and quantum correction is necessary. Bose-Einstein condensation(BEC) occur at the onset of this degeneracy. So, at the condensation temperature $\frac{kT_c}{\frac{2\pi\hbar^2}{mL^2}}\sim N^{2/3}$. However the condition of thermodynamic limit($L/\lambda_T\rightarrow\infty$) is also valid at this temperature. For finite system the thermodynamic limit is realized as $L/\lambda_T\gg 1$. However, at sufficiently ultralow temperatures $L/\lambda_T$ can be comparable to 1. In this situation, the Bose-Einstein(B-E) statistics which is valid in the thermodynamic limit needs a correction to the statistics. 

               The prescription of the derivation of B-E statistics is as follows\cite{1}. The partition function for this system of $N$ particles is $Z(N)=\sum_R e^{-(n_1\epsilon_1+n_2\epsilon_2+n_3\epsilon_3+...)/kT}$, where $\epsilon_i$ is single particle energy level and $n_i$ is the number of particles in the $i$th single particle state. For Bose gas $n_i=0,1,2,3...$ R represent all possible number distribution($\{n_i\}$) subject to the constraint $\sum_{i=0}^{\infty}n_i=N$. For this constraint the evaluation of $Z(N)$ without approximation is impossible. Since $Z(N)$ is a rapidly increasing function we choose a exponentially decreasing function $e^{\mu N/kT}$ so that $Z(N)$ is maximized at a suitable negative value of $\mu$. Thus we can approximately write $\sum_{N'}Z(N')e^{\mu N'/kT}=Z(N)e^{\mu N/kT}\triangle N'$, where $\triangle N'$ is the width of the $Z(N')$ distribution about the maximum $Z(N)$. $\mu$ is called the chemical potential. In the thermodynamic limit $N\rightarrow\infty$ and length scale $L\rightarrow\infty$. From the first condition of the thermodynamic limit we can write $\frac{\triangle N'}{N}\rightarrow 0$. Using $\frac{\triangle N'}{N}\rightarrow 0$ we can arrive at the relation $lnZ=-\mu N/kT-\sum_{i} ln (1-e^{-(\epsilon_i-\mu)/kT})$. The average no. of particles in the single particle state($i$) is $\bar{n_i}=\frac{\sum_R n_i e^{-(n_1\epsilon_1+n_2\epsilon_2+n_3\epsilon_3+...)/kT}}{\sum_R e^{-(n_1\epsilon_1+n_2\epsilon_2+n_3\epsilon_3+...)/kT}}$. Using the second condition of thermodynamic limit($L\rightarrow\infty$) we can write $\frac{\epsilon_{i+1}-\epsilon_i}{kT}\rightarrow 0$ (since $\epsilon_i\sim\frac{\hbar^2 i^2}{mL^2}$). From this condition we can write $\bar{n_i}=-kT\frac{\partial ln Z}{\partial\epsilon_i}$. So, in the thermodynamic limit we have the B-E statistics as $\bar{n_i}=\frac{1}{e^{(\epsilon_i-\mu)/kT}-1}$.

             The usual condition of thermodynamic limit is not properly satisfied in the case of the experimental setup of BEC of 3-d harmonically trapped Bose gas\cite{2,3,4,5,6}. In the semiclassical approximation the number density at the distance(r) from the center of the trap is\cite{7} $\bar{n}(r)\sim e^{-\frac{m\omega^2r^2}{2kT}}$, where $\frac{1}{2}m\omega^2r^2$ is the trap potential and $\omega$ is the angular trap frequency. The length scale of this 3-d trapped Bose gas is $L\sim \sqrt{\frac{2kT}{m\omega^2}}$. Putting this expression of length in $\frac{kT_c}{\frac{2\pi\hbar^2}{mL^2}}\gg N^{2/3}$ we get the condition of classical limit as $\frac{kT}{\hbar\omega}\gg N^{1/3}$, condition of thermodynamic limit($L/\lambda_T\rightarrow\infty$) as $\frac{kT}{\hbar\omega}\rightarrow\infty$ and we get the condensation temperature for this system as $T_c\sim\frac{\hbar\omega}{k}N^{1/3}$. More precisely the thermodynamic limit for this system is realized as $\omega\rightarrow 0$,$N\rightarrow\infty$ and $N\omega^3=constant$. Exact calculation with thermodynamic limit show that\cite{7} $L=\sqrt{\frac{\hbar}{m\omega}}\sqrt{\frac{2\zeta(4)kT}{\zeta(3)\hbar\omega}}$ and\cite{7,8} $T_c=\frac{\hbar\omega}{k}[\frac{N}{\zeta(3)}]^{1/3}$. With the consideration of thermodynamic limit, total number excited particles at $T\le T_c$ are $N_e=[\frac{kT}{\hbar\omega}]^{3}\zeta(3)$ and the number of condensed particles are $N_o=N[1-(\frac{T}{T_c})^{3}]$. In the experimental setup the typical value of\cite{9} $N$, $\omega$ and $m$ are of the order of 50000, 2645$s^{-1}$ and 52 amu. With these experimental parameters the length scale of the system at $T_c$ is $L=\sqrt{\frac{\hbar}{m\omega}}\sqrt{\frac{2\zeta(4)kT}{\zeta(3)\hbar\omega}}\sim 56\times 10^{-4}mm.$ and $\lambda_{T_c}=2.88\times 10^{-4}mm.$ and their ratio is $L/\lambda_T=32.2$. However, at sufficiently ultralow temperatures($T\sim\frac{\hbar\omega}{k}$) well below the condensation temperature, a large fraction of particle come down to the ground state and the length scale of the system becomes $\sim \sqrt{\frac{\hbar}{m\omega}}\sim 6.76\times 10^{-4}mm.$ Here we see that at ultralow low temperatures well below the condensation temperature, the thermal de Broglie wavelength becomes comparable to the system size. At these ultralow temperatures($T\sim\frac{\hbar\omega}{k}$) the usual theory of statistical mechanics of finite system is not (properly) valid. So at $ \frac{\hbar\omega}{k}\lnsim T\lesssim T_c$ we seek a ultralow temperature as well as finite size correction to B-E statistics.

               To quantify the correction arising from ultralow temperatures, let us start from Tsallis type of generalized Bose-Einstein statistics as $\bar{n_i}=\frac{1}{[1+(q-1)\frac{(\epsilon_i-\mu)}{kT}]^{\frac{1}{(q-1)}}-1}$, where $q$ is a hidden variable \cite{10,11}. As $q\rightarrow 1$, we get back B-E statistics. The relative probability that $E$ be the total energy of a system is given by Boltzmann factor $e^{-E/kT}$. In Tsallis statistics this factor is replaced by $\frac{1}{(1+(q-1)E/kT)^{1/(q-1)}}$. As $q\rightarrow 1$, we get back Boltzmann statistics. However, Tsallis statistics\cite{12} ($\frac{1}{(1+(q-1)E/kT)^{1/(q-1)}}$) is applied to equilibrium\cite{11,13,14,15} as well as nonequilibrium\cite{16,17,18} systems. Tsallis statistics is rederived as one of the superstatistics\cite{19} of a nonequilibrium system. In this theory the hidden variable($q$) of Tsallis statistics is equated with system parameter and $q$ is no longer a variable. In the theory of superstatistics\cite{19} $2/(q-1)$ is redefined as effective no. of degrees of freedom. In the theory of dynamical foundation of nonextensive statistical mechanics, this effective no. of degrees of freedom is equated as\cite{16} $(3-q)/(q-1)$. However, for finite equilibrium system we equate $q$ with a system parameter. We equate $q-1$ with $\lambda_T/L$ so that at the thermodynamic limit($L/\lambda_T\gg 1$) of finite system we can go to the usual Bose-Einstein statistics. Fig.1 shows experimental and theoretical plot of condensation fraction with temperature. In this figure we see that a negative shift of condensation fraction is necessary to satisfy the theoretical plot with the experimental data. In the theoretical plot there is finite size correction and the correction of two body interaction. We shall see that the generalized B-E statistics with $(q-1)\propto\frac{\hbar\omega}{kT}$ will give rise to a significant negative shift of condensation fraction with temperature. This significant shift along with finite size correction and the correction of two body interaction might satisfy the experimental results. So for the trapped Bose gas we shall start from the generalized statistics $\bar{n_i}=\frac{1}{[1+\frac{\hbar\omega}{\alpha kT}\frac{(\epsilon_i-\mu)}{kT}]^{\frac{\alpha kT}{\hbar\omega}}-1}$, where $\alpha$ is an arbitrary constant. We shall determine this  $\alpha$ from the experimental result. 

             For a system of Bose gas in isotropic harmonic trap single particle energy levels are $\epsilon_j = (\frac{3}{2}+j)\hbar\omega , \ (j=0,1,2,3,...)$. 
If $T_c$ be the condensate temperature then at $ T\leq T_c$ the chemical potential $\mu=\frac{3}{2}\hbar\omega $. So at $T\leq T_c$ the number of particles from the generalized B-E statistics follows as
\bea
\bar{n_j}=\frac{1}{[1+\frac{j}{\alpha t^{2}}]^{\alpha t}-1}
\eea
where $t=\frac{KT}{\hbar\omega}$. Obviously, in the thermodynamic limit of this finite system $t\gg 1$. Since for $t\sim 1$ the thermal de Broglie become comparable to the system size  this expression of $\bar{n_j}$ is not valid for $t\lesssim 1$. For a $3-d$ isotropic harmonic oscillator the density of states is $\frac{(j^{2}+3j+2)}{2}$. So total no. of particles in the excited states at $t<t_c= \frac{kT_c}{\hbar\omega}$ will be 
\bea
N_e=\sum_{j=1}^{\infty} \frac{(j^{2}+3j+2)}{2}\frac{1}{[1+\frac{j}{\alpha t^{2}}]^{\alpha t}-1}
\eea
In this summation the volume term of the density of states(i.e. the term $j^{2}/2$) will dominate over the surface term (i.e. the term $ 3j/2$). This surface term gives finite size correction. The third term contribute insignificantly to the calculation of number of excited particles. Converting the summation into integration we have 
\bea
N_e=\int_{0}^{\infty}\frac{j^{2}+3j}{2}\frac{1}{[1+\frac{j}{\alpha t^{2}}]^{\alpha t}-1}\,dj.
\eea
For $t\gg 1$ the term ${[1+\frac{j}{\alpha t^{2}}]^{\alpha t}-1}$ in the denominator of the above equation can be approximated as ${[1+\frac{j}{\alpha t^{2}}]^{\alpha t}}\approx e^{j/t}-\frac{j^{2}}{2\alpha t^{3}}e^{j/t}+\it{O}(\frac{1}{t^{5}})$.

At $t\gg 1$, considering the significant correction terms from the above equation(3) we can write 
\begin{eqnarray}
N_e&=&\int_{0}^{\infty}\frac{j^{2}}{2}\frac{1}{e^{j/t}-1}\,dj+\int_{0}^{\infty}\frac{3j}{2}\frac{1}{e^{j/t}-1}\,dj\nonumber\\&+&\frac{1}{2\alpha t^3}\int_{0}^{\infty}\frac{j^{2}}{2}\frac{e^{j/t}}{(e^{j/t}-1)^2}\,dj\nonumber\\&=& t^{3}\zeta(3)+\frac{3t^2}{2}\zeta(2)+\frac{6t^2}{\alpha}\zeta(4)
\end{eqnarray}
Till now we did not consider the particle particle interaction. To achieve BEC it is necessary to take a very dilute gas. The gas being very dilute there should be a correction term in the expression of $N_e$ due to two body scattering. This correction within Hartree-Fock(H-F) approximation has been discussed in \cite{20}. According to H-F approximation, the correction term to the above $N_e$ is $3\times 1.326\frac{a}{\sqrt{\hbar/m\omega}}N_e^{7/6}=4.93\frac{a}{\sqrt{\hbar/m\omega}}t^{7/2}$, where $a$ is the s-wave scattering length. So more corrected expression of number of excited particle at $1\lnsim t\le t_c$ would be 
\bea
N_e=t^{3}\zeta(3)+\frac{3t^2}{2}\zeta(2)+4.93\frac{a}{\sqrt{\frac{\hbar}{m\omega}}}t^{7/2}+\frac{6t^2}{\alpha}\zeta(4)
\eea
At $T=T_c$ all the particles will be in the excited states\cite{7,8,21}. So at $T=T_c$ or at $t_c=\frac{kT_c}{\hbar\omega}$ the number of excited particle will be equal to the total number of particles. So, 
\bea
N=t_c^{3}\zeta(3)+\frac{3t_c^2}{2}\zeta(2)+4.93\frac{a}{\sqrt{\frac{\hbar}{m\omega}}}t_c^{7/2}+\frac{6t_c^2}{\alpha}\zeta(4)
\eea
In the thermodynamic limit\cite{7,8} $N_e=t^{3}\zeta(3)$ and the condensation temperature $T_o$ is such that\cite{7,8} $t_o=\left[\frac{N}{\zeta (3)}\right]^{1/3}$, where $t_o=\frac{kT_o}{\hbar\omega}$. Comparing this expression of $t_o$ and $t_c$ of equation (6) we see that $t_c<t_o$ and there is a shift $\delta t_c=t_c-t_o$ of condensation temperature due to the inclusion of finite size correction term, correction term due to two body scattering and due to the ultralow temperature correction of B-E statistics. At $t\le t_c$ the fraction of number of particles in the ground state is
\begin{eqnarray}
\frac{N_o}{N}&=&\frac{N-N_e}{N}=1- [\frac{t}{t_o}]^3\nonumber\\&-&[\frac{3t^2}{2}\zeta(2)+4.932\frac{a}{\sqrt{\frac{\hbar}{m\omega}}}t^{7/2}+\frac{6t^2}{\alpha}\zeta(4)]/[t_o^{3}\zeta(3)]\nonumber\\
\end{eqnarray}
However in the thermodynamic limit the expression of this fraction would be\cite{7,8} $[\frac{N_o}{N}]_{T-L}=1-[\frac{t}{t_o}]^3$. Now, from equation(7), due to the correction terms we get the fractional change in condensation temperature as 
\begin{eqnarray}
\frac{\delta T_c}{T_o}&=&\frac{\delta t_c}{t_o}=[[[[N-[\frac{3t_c^2}{2}\zeta(2)+4.932\frac{a}{\sqrt{\frac{\hbar}{m\omega}}}t_c^{7/2}\nonumber\\&+&\frac{6t_c^2}{\alpha}\zeta(4)]]/\zeta(3)]^{1/3}-[N/\zeta(3)]^{1/3}]]/[N/\zeta(3)]^{1/3}\nonumber\\&\approx& -\frac{1}{3}[\frac{3t_c^2}{2}\zeta(2)+4.932\frac{a}{\sqrt{\frac{\hbar}{m\omega}}}t_c^{7/2}\nonumber\\&+&\frac{6t_c^2}{\alpha}\zeta(4)\zeta(3)]/N 
\end{eqnarray} 
Putting $t_c\approx [\frac{N}{\zeta(3)}]^{1/3}$ in the above equation we get
\begin{eqnarray}
\frac{\delta t_c}{t_o}=&-&\frac{\zeta(2)}{2[\zeta(3)]^{2/3}}N^{-1/3}-1.326\frac{a}{\sqrt{\hbar/m\omega}}N^{1/6}\nonumber\\&-&\frac{2\zeta(4)}{\alpha[\zeta(3)]^{2/3}}N^{-1/3}
\end{eqnarray}
From the first term of the above equation(9) we get the $T_c$ shift due to the finite size correction as \cite{22} $\frac{\delta t_{c}^{f-s}}{t_o}=-\frac{\zeta(2)}{2[\zeta(3)]^{2/3}}N^{-1/3}=-.728N^{-1/3}$ and for 50000 particles we get $\frac{\delta t_{c}^{f-s}}{t_o}=-1.97\%$. 
From the second term of the above equation(9) we get the $T_c$ shift due to the correction of two body interaction as \cite{7,20} $\frac{\delta t_{c}^{int.}}{t_o}=-1.326\frac{a}{\sqrt{\hbar/m\omega}}N^{1/6}=-6.61\%$ for $\omega=2645s^{-1}$, $a=105 a_B$ and for N=50000 \cite{9}. For this setup $t_o=[\frac{N}{\zeta(3)}]^{1/3}=34.65$. From the third term of the above equation(9) we get the $T_c$ shift due to ultralow temperature correction of B-E statistics as $\frac{\delta t_{c}^{ult}}{t_o}=-\frac{2\zeta(4)}{\alpha[\zeta(3)]^{2/3}}N^{-1/3}=-\frac{1}{\alpha}5.19\%$ for 50000 particles. According to the experiment $\frac{\delta t_c}{t_o}$ should be $10\%$ and since $\frac{\delta t_c}{t_o}=\frac{\delta t_{c}^{f-s}}{t_o}+\frac{\delta t_{c}^{int.}}{t_o}+\frac{\delta t_{c}^{ult}}{t_o}$ we get $\alpha=3.6549$ However, since these shifts are not negligible, in equation (8) we have to take higher order terms into account. As the corrections are not negligible we should not put $t_c=t_o$ in equation (9). Since $t_c<t_o$, the percentage $t_c^{f-s}$ and $t_c^{int}$ shift we calculated will be lowered. Using equation (7), from the numerical plot of $\frac{N_o}{N}$ with temperature we get $\frac{\delta t_{c}^{f-s}}{t_c}=-1.93\%$, $\frac{\delta t_{c}^{int.}}{t_c}=-5.75\%$. From equation (7), in the numerical plot $10\%$  $t_c$ shift will be achieved if we put $\alpha=1.48$. 
\begin{figure}
\includegraphics{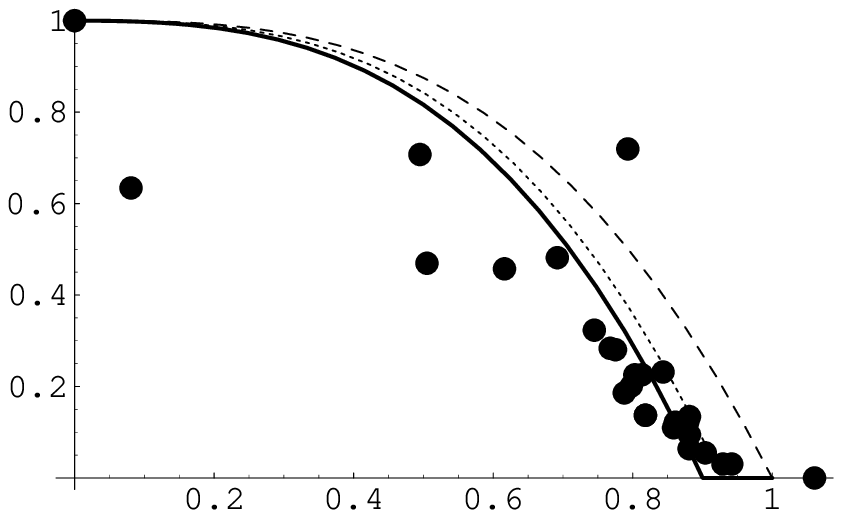}
\caption { Condensation fraction ($\frac{N_o}{N}$) {} versus  temperature ($\frac{t}{t_o}$){}plot. The thick line follows from equation(7). The dashed line corresponds to the thermodynamic limit and excludes all the correction terms in equation(7). The dotted line corresponds to the finite size correction and the correction due to interaction and exclude the ultralow temperature correction term in equation (7). All the theoretical curves are drawn according to the following experimental parameters. The dotted points are experimental points of Bose-Einstein condensation of $^{52}Cr$ where\cite{9} $N=50000$,$\omega =2645s^{-1}$, $m=52$a.m.u.,$a=105 a_B$ and $T_o\sim 700 nK$} 
\label{fig:Condensation Fraction}
\end{figure}
              Since in the experimental setup, below the condensation temperature $\frac{\lambda_{T}}{L}\sim t\lesssim 34$, we introduced corrected statistics ($n_j=\frac{1}{[1+\frac{j}{\alpha t^{2}}]^{\alpha t}-1}$) apart from B-E statistics ($n_j=\frac{1}{e^{j/t}-1})$ which is absolutely valid when $t\rightarrow\infty$. Well below $T_c$, as $t$ approaches to 1 the B-E statistics needs more corrections. Below $T_c$ as a comparative study of the two statistics, let us calculate the ratio of specific heat from the two statistics. As a comparative study we can take ideal gas and we disregard the surface term in equation (3) to get 
\begin{eqnarray}
N_e(t)=\sum_{i=1}^{\infty}(\alpha t^{2})^{3}\frac{\Gamma (\alpha it -3)}{\Gamma (\alpha it)}.
\end{eqnarray}
As $t\gg 1$,\cite{7,8} $N_e(t)\rightarrow t^{3}\zeta (3)$ as expected from B-E statistics. The total energy of the system would be 
\begin{eqnarray}
E(t) &\sim& (\hbar\omega)\int_{0}^{\infty}\frac{j^{3}}{2}\frac{1}{[1+
\frac{j}{\alpha t^{2}}]^{\alpha t}-1}\,dj \nonumber\\
&=& 3(\hbar\omega)\sum_{i=1}^{\infty}(\alpha t^{2})^{4}\frac{\Gamma (\alpha it -4)}{\Gamma (\alpha it)}
\end{eqnarray}
Now the specific heat would be 
$C_v(t)=\frac{k}{\hbar\omega}\frac{d}{dt}E(t)$
\bea
=\int_{0}^{\infty} k\frac{j^{3}(1+\frac{j}{\alpha t^{2}})^{\alpha t}\left[\frac{2j}{t^{2}(1+\frac{j}{\alpha t^{2}})}-\alpha\log [1+\frac{j}{\alpha t^{2}}]\right]}{2\left[\left(1+\frac{j}{\alpha t^{2}}\right)^{\alpha t}-1\right]^{2}}\,dj.
\eea
As $t\gg 1$, $E(t)\rightarrow 3\hbar\omega\zeta (4)t^{4}$ as expected from the B-E statistics\cite{7,8}. Let us denote the ratio of specific heat calculated from the corrected B-E statistics and from the B-E statistics as 
\bea
r'(t)=\frac{C_v}{k12\zeta(4)t^{3}} 
\eea
Obviously as $t\rightarrow \infty$, $r'\rightarrow 1$. With $\alpha=1.48$ numerical plot of $r(t)$  at very low temperatures ($1\lnsim t\le t_c$) is shown in the FIG.2. 
\begin{figure}
\includegraphics{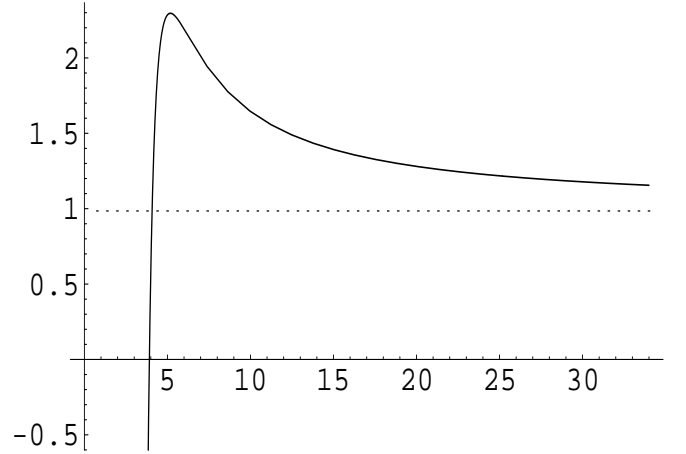}
\caption { ($r'(t) =\frac{C_v}{k12\zeta (4) t^{3}}$) {} and  Temperature ($t=\frac{KT}{\hbar\omega}$){},where $C_v$ follows from equation(12).The doted line shows the theoretical value of the ratio $r'(t)$ in thermodynamic limit when $\omega\rightarrow 0$.} 
\label{fig:RATIO OF SPECIFIC HEATS}
\end{figure}        
         In  FIG. 2 we see that at high temperature $t\gg 1$, the specific heat behaves well according to our familiar $T^{3}$ law. At $to=34.65$, the numerical value of $r'$ is 1.15 and at $t=15$, the numerical value of $r'$ is 1.39. So at $t=15$ there is $39\%$ error in the specific heat. In this figure we see that this specific heat law dose not holds fairly well at $t\lesssim 6$. From this figure we also see that to achieve the thermodynamical limit, $\omega\rightarrow 0$ is no longer a necessary criterion. If $\omega\sim\frac{kT}{15\hbar}$, the system (3-d isotropic harmonically trapped Bose gas) will correspond to thermodynamic limit with $\sim 39\%$ error in the specific heat. Similarly we can say that to achieve the thermodynamic limit of a system of particles in a box the volume need not necessarily tend to infinity. In this way we can also estimate the minimum volume of a thermodynamical system. In the FIG. 2 we see that specific heat for this finite system becomes negative at $t\lesssim 5$. At this range of temperatures $L\lesssim\lambda_T$ and the theory of statistical mechanics is not valid. In spite of that, due to nonextensivity the appearance of negative specific at these temperatures is not surprising theoretically\cite{23} and experimentally\cite{24}.  
\begin{figure}
\includegraphics{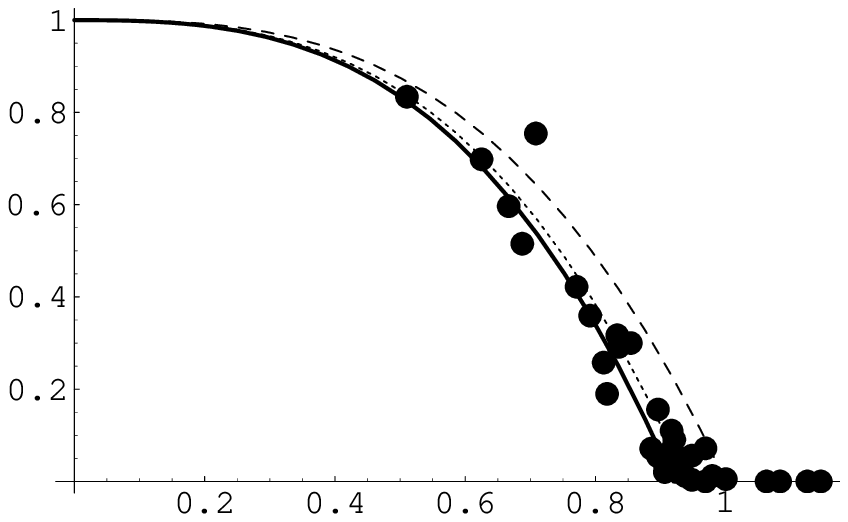}
\caption { Condensation fraction ($\frac{N_o}{N}$) {} versus  temperature ($\frac{t}{t_o}$){}plot. The thick line follows from equation(7). The dashed line corresponds to the thermodynamic limit and excludes all the correction terms in equation(7). The dotted line corresponds to the finite size correction and the correction due to interaction and exclude the ultralow temperature correction term in equation (7). All the theoretical curves are drawn according to the following experimental parameters. The dotted points are experimental points of Bose-Einstein condensation of $^{87}Rb$ where\cite{25} $N=40000$,$\omega =1140s^{-1}$, $m=87$a.m.u.,$a=90a_B$ and $T_o=280 nK$} 
\label{fig:Condensation Fraction2}
\end{figure}
            For smaller $\omega$ the system size($L=\sqrt{\frac{\hbar}{m\omega}}\sqrt{\frac{2\zeta(4)kT}{\zeta(3)\hbar\omega}}$) is larger and $\frac{\lambda_T}{L}$ is smaller. In this case the the ultralow temperature correction would be smaller. In the experiment \cite{25} $\omega=1140s^{-1}$ smaller than that of the previous case. For this experimental setup the numerical value of $\frac{\delta t_{c}^{f-s}}{t_o}=-2.1\%$ and $\frac{\delta t_{c}^{int.}}{t_o}=-5.5\%$. From the experimental data as shown in FIG 3, we see that $\frac{\delta t_{c}}{t_o}\sim 8\%$. So $\frac{\delta t_{c}^{int}}{t_o}=-.4\%$ and it corresponds to $\alpha=5.5$. Due to the larger system size the ultralow temperature correction is smaller as we see in the FIG 3.
           
           In the thermodynamic limit of a finite system $L/\lambda_T\gg 1$. Bose-Einstein statistics is absolutely valid at this limit. However for trapped Bose gas at $T\lesssim T_c$ this ratio is\cite{9} $\sim 32$ and at temperatures well below $T_c$ this ratio is comparable to 1. For this reason we expect a correction to B-E statistics. The correction to the statistics depends on this ratio. As a correction we introduce a nonextensive type of statistics ($\bar{n_i}=\frac{1}{[1+\frac{\hbar\omega}{\alpha kT}\frac{(\epsilon_i-\mu)}{kT}]^{\frac{\alpha kT}{\hbar\omega}}-1}$). Smaller the ratio larger is the correction and it is evident in FIG 1 and in FIG 3 of the BEC experiments. Although the experimental uncertainties are larger with respect to these theoretical corrections yet the ultralow temperature correction might be of our interest. However, this ultralow temperature correction is valid only for finite system. How $\alpha$ is to be determined theoretically remains an open question.         
          
           Several useful discussions with J.K. Bhattacharjee and with Koushik Ray of I.A.C.S. are gratefully acknowledged.

\end{document}